\documentclass[12pt,onecolumn,journal]{IEEEtran}
\usepackage{latexsym}
\usepackage{amsfonts}
\usepackage{amsbsy}
\usepackage{amsmath,amssymb}
\usepackage{times}
\usepackage{graphicx}
\usepackage{stkernel}
\usepackage{enumerate}
\usepackage[usenames]{color}
\usepackage[dvips]{pstcol}
\usepackage{epstopdf}
\usepackage{amsfonts,amssymb,amsmath,bm}
\usepackage{graphicx,subfigure}
\usepackage{epstopdf}
\usepackage{cite,color}
\usepackage{stfloats}
\usepackage{enumerate}
\usepackage{algorithm}
\usepackage{cite}
\usepackage[justification=raggedright,font={small}]{caption}

\input epsf

\newtheorem{theorem}{\textbf{Theorem}}
\newtheorem{lemma}{\textbf{Lemma}}

\newtheorem{assump}{\textbf{Assumption}}

\newcommand{\trace}{{\rm Tr}}

\newcommand{\bI}{\mathbf{I}}
\newcommand{\bH}{\mathbf{H}}

\newcommand{\bF}{\mathbf{F}}
\newcommand{\bQ}{\mathbf{Q}}

\newcommand{\bA}{\mathbf{A}}
\newcommand{\bB}{\mathbf{B}}
\newcommand{\bC}{\mathbf{C}}

\newcommand{\bU}{\mathbf{U}}
\newcommand{\bV}{\mathbf{V}}

\newcommand{\bx}{\bm{x}}

\newcommand{\by}{\bm{y}}

\newcommand{\hs}{\hat{\mathbf{s}}}

\newcommand{\bu}{\bm{u}}

\newcommand{\bs}{\bm{s}}

\newcommand{\bn}{\bm{n}}
\newcommand{\Cdom}{\mathbb{C}}

\newcommand{\cgauss}{\mathcal{CN}}
\newcommand{\opmin}{\mathop{\mathrm{minimize}}\limits}
\newcommand{\opmax}{\mathop{\mathrm{maximize}}\limits}
\ifCLASSINFOpdf
\else
\fi
\begin{document}
\title{Hybrid Beamforming for Massive MIMO Over-the-Air Computation}
\author{Xiongfei Zhai,
        Xihan Chen,
        Jie Xu,
        and Derrick Wing Kwan Ng
        \thanks{
X. Zhai is with the School of Information Engineering, Guangdong University of Technology, Guangzhou 510006, China (e-mail: zhaixiongfei@gdut.edu.cn).

X. Chen is with the College of Information Science and Electronic Engineering, Zhejiang University, Hangzhou 310000, China (e-mail: chenxihan@zju.edu.cn).

J. Xu is with the Future Network of Intelligence Institute (FNii) and the School of Science and Engineering, The Chinese University of Hong Kong, Shenzhen, Shenzhen 518172, China (e-mail: xujie@cuhk.edu.cn). J. Xu is the corresponding author.

D. W. K. Ng is with the School of Electrical Engineering and Telecommunications, University of New South Wales, Sydney, NSW 2052, Australia (email: w.k.ng@unsw.edu.au).
}
}
\maketitle
\begin{abstract}
Over-the-air computation (AirComp) has been recognized as a promising technique in Internet-of-Things (IoT) networks for fast data aggregation from a large number of wireless devices. However, as the number of devices becomes large, the computational accuracy of AirComp would seriously degrade due to the vanishing signal-to-noise ratio (SNR). To address this issue, we exploit the massive multiple-input multiple-output (MIMO) with hybrid beamforming, in order to enhance the computational accuracy of AirComp in a cost-effective manner. In particular, we consider the scenario with a large number of multi-antenna devices simultaneously sending data to an access point (AP) equipped with massive antennas for functional computation over the air. Under this setup, we jointly optimize the transmit digital beamforming at the wireless devices and the receive hybrid beamforming at the AP, with the objective of minimizing the computational mean-squared error (MSE) subject to the individual transmit power constraints at the wireless devices. To solve the non-convex hybrid beamforming design optimization problem, we propose an alternating-optimization-based approach, in which the transmit digital beamforming and the receive analog and digital beamforming are optimized in an alternating manner. In particular, we propose two computationally efficient algorithms to handle the challenging receive analog beamforming problem, by exploiting the techniques of successive convex approximation (SCA) and block coordinate descent (BCD), respectively. It is shown that for the special case with a fully-digital receiver at the AP, the achieved MSE of the massive MIMO AirComp system is inversely proportional to the number of receive antennas. Furthermore, numerical results show that the proposed hybrid beamforming design substantially enhances the computation MSE performance as compared to other benchmark schemes, while the SCA-based algorithm performs closely to the performance upper bound achieved by the fully-digital beamforming.
\end{abstract}
\IEEEpeerreviewmaketitle
\section{Introduction}
Future Internet-of-Things (IoT) networks need to support an enormous number of wireless devices for sensing the environment, aggregate massive sensing data for analysis, and accordingly take physical actions\cite{Agiwal2016,Lin2017}. Conventionally, such data aggregation is implemented via wireless devices individually sending their data to an access point (AP) or a fusion center over wireless multiple access channels, in which the AP may need to decode the individual messages from each device by treating messages from others as harmful interference. Nevertheless, for practical IoT applications, the AP may be interested in computing a certain function value (e.g., the sum value) of the aggregated data rather than the individual messages (e.g., in federated learning setups \cite{Yang2020}). In this case, the above conventional multiple access scheme may not be energy- or spectral-efficient and may also lead to excessively long network latency, especially when the number of devices becomes significantly large. To overcome the drawback, the over-the-air computation (AirComp) technique has been proposed recently, which utilizes the co-channel interference among devices as a beneficial factor for functional computation \cite{Abari2016,ZhuAir1}. By exploiting the signal superposition property of multiple access channels, the AirComp technique is able to directly compute a class of nomographic functions (e.g., arithmetic mean, weighted sum, geometric mean, polynomial, and Euclidean norm) of distributed sensing data from the concurrent transmission of distributed wireless devices \cite{Boche2015}.



In general, AirComp can be implemented in both analog and digital modes. While the simple uncoded analog transmission was shown to achieve the minimum functional distortion when the data sources follow the independent and identically distributed (i.i.d.) Gaussian distribution \cite{Gastpar2008}, coding was shown to be  necessary for improving the computation performance under the bivariate Gaussian \cite{Wagner2008} and correlated Gaussian \cite{Soundararajan2012} distributed data sources. Besides, for analog AirComp, the computation mean-squared error (MSE) is normally adopted as the performance metric. In the single-antenna setup, a proper power control is essential for minimizing the computation MSE \cite{Xiao2008,Wang2011,Cao2019}. For instance, under the coherent multiple access channel, the optimal power control strategy for minimizing the computation MSE was proposed in \cite{Xiao2008} by using convex optimization, and optimal power allocation strategies were proposed in \cite{Wang2011} for minimizing the distortion outage probability (defined as the probability that the computation MSE exceeds a given threshold). Furthermore, under fading channels, the optimal power allocation strategy for minimizing the average computation MSE was studied \cite{Cao2019}. In particular, multi-antenna beamforming is an efficient technique to further enhance the computation MSE performance. For instance, the authors in \cite{ZhuAir1} investigated the multiple-input multiple-output (MIMO) AirComp for computing multiple functions simultaneously, in which a closed-form equalization at the AP was proposed to minimize the computation MSE under the zero-forcing (ZF) transmit beamforming at wireless devices. Notice that the implementation of AirComp requires the synchronization among all devices; towards this end, the so-called AirShare design was developed in \cite{Abari2015}, where a shared clock was broadcast to all devices to facilitate synchronization.

On the other hand, digital AirComp was proposed to enhance computational accuracy via proper coding methods \cite{Nazer2007,Zhang2006,Appuswamy2014,Erez2005,Nazer2011,Jeon2016}. The idea of digital AirComp first appeared for functional computation in wireless sensor networks (see, e.g., \cite{Nazer2007}) and physical-layer network coding in a two-way relay channel (see, e.g., \cite{Zhang2006}). For AirComp, the achievable computation rate under different system setups was characterized in \cite{Nazer2011} and \cite{Jeon2016}, which is defined as the number of functional values computed per unit time under a predefined computational accuracy. Furthermore, to enable multi-function computation with enhanced computation rate, the authors in \cite{Wu2020} integrated the idea of non-orthogonal multiple access (NOMA) \cite{Dai2015} in AirComp, in which multiple functions from different wireless devices are superposed in each resource block.



In this paper, we particularly focus our study on the analog AirComp, in which wireless devices send uncoded data to a single AP for functional computation. In practice, the implementation of AirComp over large-scale wireless networks faces several technical challenges. For instance, as the number of wireless devices increases, the computation performance in AirComp systems may seriously degrade, due to the vanishing of the received signal-to-noise ratio (SNR) at the AP. As a remedy, massive MIMO \cite{Marzetta2010} is considered as a promising viable solution to improve the computational accuracy of AirComp systems by exploiting its rich spatial degrees of freedom and tremendous array gain. To the best of our knowledge, the exploitation of massive MIMO for AirComp has not been reported in the literature yet. Hence, it motivates us to consider the combination of massive MIMO and AirComp to improve the computation accuracy.

Despite the potential benefits, the amalgamation of massive MIMO and AirComp would incur high fabrication cost and energy consumption due to the conditional large numbers of radio frequency (RF) chains as well as analog-to-digital converters (ADCs) associated with antennas. To address these issues, the hybrid beamforming structure has emerged as a promising solution, which allows the AP to employ a smaller number of RF chains than that of antenna elements \cite{Sudarshan2006,Rial2016,Yang2018,Hur2013,Alkhateeb2014,Han2019,Zhai2017,Yu2016,Dsouza2018,Liu2019,Song2020}. Considering the tradeoff between performance versus complexity, different types of hybrid beamforming (e.g., fully-connected \cite{Hur2013,Alkhateeb2014,Yu2016,Zhai2017,Yang2018,Han2019,Liu2019} and partially-connected \cite{Sudarshan2006,Rial2016,Dsouza2018,Song2020}) with different kinds of RF electronic modules (e.g., analog switches \cite{Sudarshan2006,Rial2016} and analog phase shifters \cite{Hur2013,Alkhateeb2014,Yu2016,Zhai2017,Yang2018,Dsouza2018,Han2019,Liu2019,Song2020}) have been investigated. In particular, the fully-connected hybrid beamforming is appealing as it can achieve better performance with each RF chain connected to all antennas, while the partially-connected hybrid beamforming shows a lower complexity but compromised performance, since it allows every RF chain to connect to only part of antennas.


Although there have been a handful of prior works investigating hybrid beamforming for massive MIMO communication systems \cite{Sudarshan2006,Rial2016,Yang2018,Hur2013,Alkhateeb2014,Han2019,Zhai2017,Yu2016,Dsouza2018,Liu2019,Song2020}, these designs cannot be directly applied to massive MIMO AirComp systems, due to the following reasons. First, the design objectives are fundamentally different. In massive MIMO communication systems, the hybrid beamforming design aims to maximize the communication rate by eliminating the interference, while in massive MIMO AirComp systems, the hybrid beamforming design targets for minimizing the computation error by harnessing the ``interference". Second, inspired by the low-latency requirement of data-intensive IoT applications, the hybrid beamforming in massive MIMO AirComp systems should achieve good computational accuracy with a less computational complexity/latency. Motivated by the above observations, in this paper, we investigate the hybrid beamforming design for massive MIMO AirComp systems. As an initial attempt, we adopt the fully-connected hybrid beamforming with analog phase shifters to achieve full spatial degrees of freedom of massive MIMO for AirComp.

The main results of this paper are summarized as follows.

\begin{itemize}
\item \; We consider a massive MIMO AirComp system with a massive-antenna AP and a massive number of wireless devices, which aims to compute multiple arithmetic sum functions of the recorded signals from all the devices. Our objective is to jointly optimize the transmit digital beamforming, the receive analog beamforming, and the receive digital beamforming to minimize the computation MSE. The formulated problem is generally intractable due to the highly coupled variables in the objective function and the constant modulus constraints for the receive analog beamforming.
\item \; To address the non-convex hybrid beamforming optimization problem, we propose an alternating-optimization-based approach to alternately optimize the transmit beamforming and the receive analog and digital beamforming. In particular, we optimize the receive analog beamforming by applying two methods, namely the successive convex approximation (SCA) and block coordinate descent (BCD), respectively. While the SCA-based method leads to a better performance, the BCD-based method enjoys a lower computational complexity at the expense of a compromised performance.
\item \; To gain more insights, we analyze the MSE performance for a special case with a fully-digital receiver at the AP. In this case, we show that the optimal (digital) receive beamforming follows the sum-minimization-MSE (sum-MMSE) structure. With the help of the sum-MMSE receive beamforming, we prove that the computation MSE is inversely proportional to the number of receive antennas when the AP adopts the techniques of massive MIMO.
\item \; Furthermore, numerical results show that the proposed hybrid beamforming design substantially improves the computation MSE performance of multi-function/multi-modal massive MIMO AirComp systems as compared to other benchmark scheme inspired by the ZF-based fully-digital beamforming design in \cite{ZhuAir1}. The SCA-based algorithm is shown to perform closely to the performance upper bound achieved by the fully-digital beamforming.
\end{itemize}

%
%

The remainder of this paper is organized as follows. Section \uppercase\expandafter{\romannumeral2} introduces the system model and formulates the computational MSE minimization problem. Section \uppercase\expandafter{\romannumeral3} presents the alternating-optimization-based approaches to address the formulated problem, where two algorithms are proposed to optimize the receive analog beamforming based on SCA and BCD, respectively. Section \uppercase\expandafter{\romannumeral4} analyzes the MSE performance for the special case with fully-digital beamforming design. Section \uppercase\expandafter{\romannumeral5} presents the numerical results. Finally, Section \uppercase\expandafter{\romannumeral6} draws the conclusion.

{\it{Notations}}: Throughout this paper, we adopt bold upper-case letters for matrices and bold lower-case letters for vectors. For a matrix $\bA$, $\bA(i,j)$ represents the entry on the $i^{th}$ row and the $j^{th}$ column, while $\bA^\dag$, $\bA^{*}$, $\bA^T$, and $\bA^H$ denote its Moore-Penrose pseudo inverse, conjugate, transpose, and Hermitian transpose, respectively. Furthermore, $\bI$ denotes the identity matrix whose dimension will be clear from the context, and $\Cdom^{m\times n}$ denotes the $m$-by-$n$ dimensional complex space. The notations $\mathbb{E}(\cdot)$, $\trace(\cdot)$, $\det(\cdot)$, $\text{vec}(\cdot)$, $\mathbb{R}(\cdot)$, and $\|\cdot\|$ represent the expectation, trace, determinant, vectorization, real part, and Frobenius norm of an input variable, respectively. $\bigtriangledown_{\bx} f(\bx)$ denotes the gradient of $f(\bx)$ with respect to $\bx$. $\circ$ is the Hadamard product between two matrices. The circularly symmetric complex Gaussian (CSCG) distribution with mean $\bm{\Upsilon}$ and covariance matrix $\bm{\Phi}$ is denoted by $\cgauss(\bm{\Upsilon},\bm{\Phi})$. $\text{mod}(a,b)$ is the modulus operation of $a$ with respect to $b$.

\section{System Model And Problem Formulation}
\subsection{System Model}
\begin{figure}[htbp]
\centering
\includegraphics[width=5in]{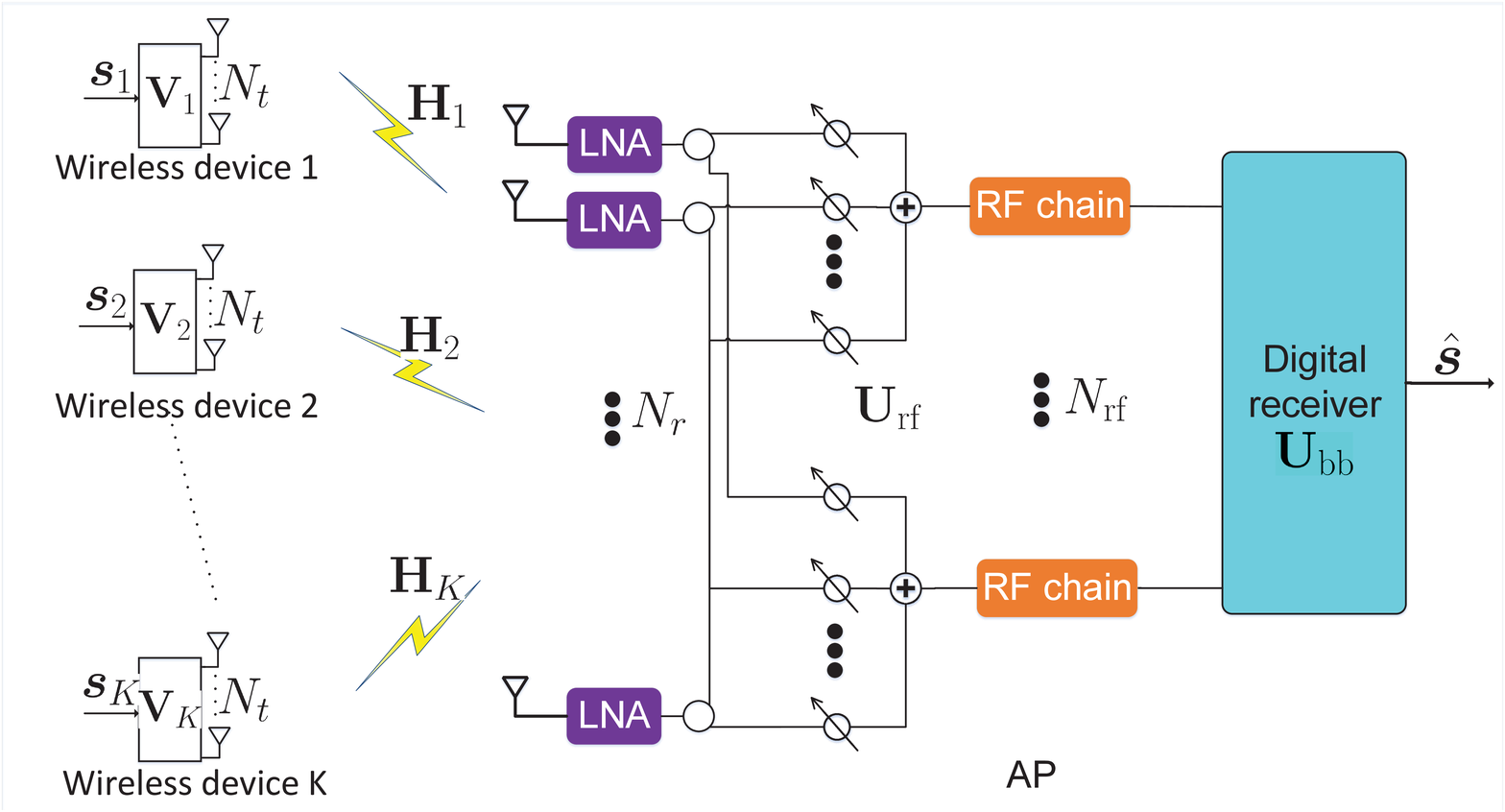}
\captionsetup{justification=raggedright}
\caption{A massive MIMO AirComp system with hybrid beamforming.}
\label{fig:fig2}
\end{figure}
We consider a massive MIMO AirComp system as shown in Fig. \ref{fig:fig2}, where the AP simultaneously serves $K$ wireless devices. Suppose each device is equipped with $N_t$ transmit antennas and the AP is equipped with $N_r$ receive antenna elements, each of which is connected to a low-noise amplifier (LNA). With massive MIMO, it is assumed that $N_r\gg N_t$. Besides, fully-digital beamforming is adopted at the wireless devices while hybrid beamforming with only $N_{\mathrm{rf}}$ RF chains is implemented at the AP, with $N_{\mathrm{rf}}\leq N_r$, to reduce the hardware cost and implementation complexity \cite{Zhai2017,Yu2016}. For the purpose of initial investigation, we consider the fully-connected hybrid beamforming with phase shifters to achieve full spatial degrees of freedom of massive MIMO. For the convenience of expression, we denote $\mathcal{N}_{r}\triangleq \{1,2,\ldots,N_{r}\}$ and $\mathcal{N}_{\mathrm{rf}}\triangleq \{1,2,\ldots,N_{\mathrm{rf}}\}$ as the antenna and RF chain sets, respectively.

In this massive MIMO AirComp system, every device records $L$ heterogeneous time-varying parameters (e.g., humidity, temperature, noise) of the environment with $L\leq \min(N_{\mathrm{rf}},N_t)$. In particular, since we focus on studying multi-function/multi-modal massive MIMO AirComp systems, we have $L>1$. The devices simultaneously transmit the recorded data to the AP for computation. At a particular time slot, we denote $s_{kl}$ as the recorded data of the $l$th parameter at device $k$, $k\in\mathcal{K}\triangleq \{1,2,\ldots,K\}, l\in\mathcal{L}\triangleq\{1,2,\ldots,L\}$, and $\bs_k=[s_{k1},\cdots,s_{KL}]^T\in\Cdom^{L\times 1}$ as the composite record vector at that device. Without loss of generality, the collected data vector is assumed to be normalized and independent form each other, i.e., $\mathbb{E}(\bs_k\bs_k^H)=\bI, \mathbb{E}(\bs_k\bs_j^H)=\bm{0}, \forall k,j\in\mathcal{K},k\neq j$, where the normalization factor for each data type is uniform for all devices and can be inverted at the AP for recovering the original data.

 In order to support ultrafast data computation, the AP exploits the superposition property of the multiple access channel to directly compute the target nomographic function with reduced communication overheads. In this paper, we are interested in the sum operation, while the design is also extendable for other nomographic functions \cite{Boche2015}. Towards this end, each device transmits a vector $\bs_k$ and the AP is interested in estimating $\bs=\sum\limits_{k=1}^K \bs_k$, which is referred to as the target-function vector \cite{ZhuAir1}.

Referring to the proposed system model, the transmitted signal by device $k$ is denoted by
\begin{equation}
\bx_k = \bV_k\bs_k,
\end{equation}
where $\bV_k \in \Cdom^{N_t\times L}$ denotes the transmit beamforming matrix. Let $P$ denote the maximum transmit power at each device. Accordingly, we have $\mathbb{E}(\|\bx_k\|^2) = \trace(\bV_k \bV_k^H) \leq P,\forall k\in\mathcal{K}$.

It is assumed that the channel state information (CSI) is perfectly known at both the AP and the devices\footnote{Practical channel estimation methods, such as random vector quantization codebook training, limited feedback, and over-the-air signaling procedure, have been proposed in \cite{Rao2014,Soltanalian2017,Ahmed2015,Goldenbaum2014}. Assuming time division duplexing protocol, both the devices and the AP can obtain the channel by applying the above channel estimation methods and exploiting the uplink-downlink channel reciprocity.}. Then the received signal vector at the AP is given by
\begin{equation}
\by = \sum\limits_{k=1}^K\bH_k\bV_k\bs_k+\bn,
\end{equation}
where $\bH_k\in \Cdom^{N_r\times N_t}$ denotes the channel matrix from device $k$ to the AP and $\bn\in\Cdom^{N_r\times 1}$ is the AWGN vector with $\bn\sim\cgauss(\mathbf{0},\sigma^2\bI)$.

Next, the AP adopts the hybrid beamforming for AirComp. Here, the hybrid beamforming needs to be properly designed for not only harnessing part of the inter-device interference to facilitate the computation, but also eliminating the inter-function interference. Let $\bU_{\mathrm{rf}}\in\Cdom^{N_r\times N_{\mathrm{rf}}}$ denote the receive analog beamforming, whose entries have constant modulus, i.e., $|\bU_{\mathrm{rf}}(i,j)|=1$, $\forall i\in\mathcal{N}_{r}$, $j\in\mathcal{N}_{\mathrm{rf}}$, and $\bU_{\mathrm{bb}}\in \Cdom^{N_{\mathrm{rf}}\times L}$ denote the low-dimension receive digital beamforming. Therefore, the processed signal after the adopted hybrid beamforming can be expressed as
\begin{equation}\label{estimasig}
\hs=\bU_{\mathrm{bb}}^H\bU_{\mathrm{rf}}^H\by.
\end{equation}
Consequently, the computational accuracy is measured by the MSE between $\hat{\bs}$ and $\bs$, which is given by \cite{ZhuAir1}
\begin{equation}\label{MSETRUE}
\begin{split}
&\text{MSE}(\{\bV_k\},\bU_{\mathrm{rf}},\bU_{\mathrm{bb}})\\
&=\mathbb{E}[\|\bs-\hat{\bs}\|^2]\\
&=\sum\limits_{k=1}^K\trace\left[(\bU_{\mathrm{bb}}^H\bU_{\mathrm{rf}}^H\bH_k\bV_k-\bI)(\bU_{\mathrm{bb}}^H\bU_{\mathrm{rf}}^H\bH_k\bV_k-\bI)^H\right]+\sigma^2\trace(\bU_{\mathrm{bb}}^H\bU_{\mathrm{rf}}^H\bU_{\mathrm{rf}}\bU_{\mathrm{bb}}).
\end{split}
\end{equation}


\subsection{Problem Formulation}
In this work, we are interested in minimizing the MSE  defined in \eqref{MSETRUE} by jointly optimizing the transmit beamforming $\{\bV_k\}$ at the devices and the receive hybrid beamforming $\bU_{\mathrm{rf}}$ and $\bU_{\mathrm{bb}}$ at the AP, subject to the constant modulus constraints on $\bU_{\mathrm{rf}}$ and the maximum power budget constraints on $\{\bV_k\}$. In particular, the MSE minimization problem can be formulated as
\begin{equation}\label{Oripro}
\begin{split}
\mathcal{P}1\!:\;&\opmin_{\bV_k, \bU_{\mathrm{rf}},\bU_{\mathrm{bb}}} ~\text{MSE}(\{\bV_k\},\bU_{\mathrm{rf}},\bU_{\mathrm{bb}})\\
&\text{subject to} ~~\trace(\bV_k\bV_k^H)\leq P, \forall k\in\mathcal{K}\\
&~~~~~~~~~~~~~~ |\bU_{\mathrm{rf}}(i,j)|=1, \forall i\in\mathcal{N}_{r}, j\in\mathcal{N}_{\mathrm{rf}}.
\end{split}
\end{equation}

Problem $\mathcal{P}1$ is difficult to solve, as the optimization variables $\{\bV_k\}$, $\bU_{\mathrm{rf}}$, and $\bU_{\mathrm{bb}}$ are highly coupled in the objective function while the unit modulus constraints of $\bU_{\mathrm{rf}}$ are highly non-convex. Furthermore, problem $\mathcal{P}1$ aims to minimize the computation MSE for recovering $\bs=\sum\limits_{k=1}^K\bs_k$ by exploiting the interference from various wireless devices. Note that this is significantly different from the conventional hybrid beamforming design problems in massive MIMO systems which mainly maximize the communication rate (for decoding $\bs_k$'s individually) by eliminating the inter-device interference. Hence, the conventional designs are not directly applicable to the considered problem $\mathcal{P}1$. Besides, due to the requirement of ultrafast computation for data aggregation, attaining an efficient solution to problem $\mathcal{P}1$ with low computational complexity is also desirable. To the best of our knowledge, however, there lacks computationally efficient and systematic algorithms to solve such non-convex problems optimally. As a compromise approach, in the next section, by exploiting the structure of problem, we propose an alternating-optimization-based method to iteratively optimize $\{\bV_{k}\}$, $\bU_{\mathrm{rf}}$, and $\bU_{\mathrm{bb}}$.


\section{Hybrid Beamforming Designs}
In this section, a novel hybrid beamforming approach is proposed to handle problem $\mathcal{P}1$, by updating $\{\bV_k\}$, $\bU_{\mathrm{rf}}$, and $\bU_{\mathrm{bb}}$ in an alternating manner. In the following, we first optimize the transmit beamforming $\{\bV_k\}$ by using the Lagrange duality method, then we update the receive analog beamforming $\bU_{\mathrm{rf}}$ by using the techniques of SCA or BCD, and finally optimize $\bU_{\mathrm{bb}}$ by exploiting the first order optimality condition.

\subsection{Optimization of Transmit Beamforming $\{\bV_k\}$}
First, we focus on the optimization of $\{\bV_k\}$ under given $\bU_{\mathrm{rf}}$ and $\bU_{\mathrm{bb}}$. In this case, problem $\mathcal{P}1$ can be equivalently decomposed into the following $K$ subproblems each for one device $k\in\mathcal{K}$, by ignoring the constant term $\sigma^2\trace(\bU_{\mathrm{rf}}\bU_{\mathrm{bb}}\bU_{\mathrm{bb}}^H\bU_{\mathrm{rf}}^H)$:
\begin{equation}\label{ProVkli}
\begin{split}
\mathcal{P}2\!:\;&\opmin_{\bV_k} \trace\left[(\bU_{\mathrm{bb}}^H\bU_{\mathrm{rf}}^H\bH_k\bV_k-\bI)(\bU_{\mathrm{bb}}^H\bU_{\mathrm{rf}}^H\bH_k\bV_k-\bI)^H\right]\\
&\text{subject to}~\trace(\bV_k\bV_k^H)\leq P, \forall k\in\mathcal{K}.
\end{split}
\end{equation}
Problem $\mathcal{P}2$ is a convex quadratic optimization problem that satisfies the Slater's constraint condition, and therefore, this problem can be optimally solved by using standard convex optimization techniques \cite{Boyd2004}. To gain more insights, we apply the Lagrange duality method to find a semi-closed-form optimal solution, which is summarized in the following lemma.
\lemma The optimal transmit beamforming solution to problem $\mathcal{P}2$ for device $k$ is given by:
\begin{equation}\label{Vkoptli}
\bV_k^{\mathrm{opt}} = (\bH_k^H\bU_{\mathrm{rf}}\bU_{\mathrm{bb}}\bU_{\mathrm{bb}}^H\bU_{\mathrm{rf}}^H\bH_k+\mu_k^{\mathrm{opt}}\bI)^{-1}\bH_k^H\bU_{\mathrm{rf}}\bU_{\mathrm{bb}},
\end{equation}
where $\mu_k^{\mathrm{opt}}\geq 0, k\in\mathcal{K},$ denotes the optimal Lagrange multiplier associated with the power constraint for device $k$ in problem $\mathcal{P}2$. Here, if $\bH_k^H\bU_{\mathrm{rf}}\bU_{\mathrm{bb}}\bU_{\mathrm{bb}}^H\bU_{\mathrm{rf}}^H\bH_k$ is invertible and
\begin{equation}\label{mu0}
\trace((\bH_k^H\bU_{\mathrm{rf}}\bU_{\mathrm{bb}}\bU_{\mathrm{bb}}^H\bU_{\mathrm{rf}}^H\bH_k)^{-2}\bH_k^H\bU_{\mathrm{rf}}\bU_{\mathrm{bb}}\bU_{\mathrm{bb}}^H\bU_{\mathrm{rf}}^H\bH_k)<P
\end{equation}
holds, we have $\mu_k^{\mathrm{opt}}=0$; otherwise, $\mu_k^{\mathrm{opt}}$ is chosen such that the equality in \eqref{mu1} holds.
\begin{equation}\label{mu1}
\trace((\bH_k^H\bU_{\mathrm{rf}}\bU_{\mathrm{bb}}\bU_{\mathrm{bb}}^H\bU_{\mathrm{rf}}^H\bH_k+\mu_k^{\mathrm{opt}}\bI)^{-2}\bH_k^H\bU_{\mathrm{rf}}\bU_{\mathrm{bb}}\bU_{\mathrm{bb}}^H\bU_{\mathrm{rf}}^H\bH_k)= P.
\end{equation}

\begin{IEEEproof}
See Appendix A.
\end{IEEEproof}

From Lemma 1, we can see that $\{\bV_k\}$ is optimized by considering the following two cases. If the transmit power budget $P$ is sufficiently large, then we choose $\mu_k^{\mathrm{opt}}=0$ such that the objective function value of problem $\mathcal{P}2$ is forced to be zero; otherwise, if the transmit power budget $P$ is limited, then we choose $\mu_k^{\mathrm{opt}}$ such that the transmit power is fully used to minimize the computation MSE.


\subsection{Optimization of Receive Analog Beamforming $\bU_{\mathrm{rf}}$}
In this subsection, we optimize $\bU_{\mathrm{rf}}$ under given $\{\bV_k\}$ and $\bU_{\mathrm{bb}}$, for which the problem is given by
\begin{equation}\label{ProUrfli}
\begin{split}
\mathcal{P}3\!:\;&\opmin_{\bU_{\mathrm{rf}}} \text{MSE}(\{\bV_k\},\bU_{\mathrm{rf}},\bU_{\mathrm{bb}})\\
&\text{subject to} ~~|\bU_{\mathrm{rf}}(i,j)|=1, \forall i\in\mathcal{N}_{r}, j\in\mathcal{N}_{\mathrm{rf}}.
\end{split}
\end{equation}
Problem $\mathcal{P}3$ is still challenging to solve mainly due to the constant modulus constraints which are intrinsically non-convex. To address this issue, we propose two algorithms by using SCA and BCD, respectively.
\subsubsection{SCA}
To gain more insights, motivated by \cite{Liu2019}, we transform problem $\mathcal{P}3$ into a more tractable form by exploiting the SCA method. To start with, we first rewrite the constant modulus constraints in its exponential form. Let $\bu_{\mathrm{rf}}=\text{vec}(\bU_{\mathrm{rf}})\in\Cdom^{N_rN_{\mathrm{rf}}\times 1}$ and $\bm{\theta}\triangleq [\theta_1,\theta_2,\ldots,\theta_{N_rN_{\mathrm{rf}}}]^T$ denote the vectorization of $\bU_{\mathrm{rf}}$ and the corresponding phase vector of $\bu_{\mathrm{rf}}$, respectively. Then problem $\mathcal{P}3$ is transformed to the following equivalent problem:
\begin{equation}\label{Prothetavec}
\begin{split}
\mathcal{P}4\!:\;&\opmin_{\bm{\theta}} f(\bm{\theta})\\
&\text{subject to} ~~ -\pi\leq \bm{\theta}(i)\leq \pi, \forall i\in\mathcal{Y},
\end{split}
\end{equation}
where
\begin{align}
&f(\bm{\theta})=\text{MSE}(\{\bV_k\},\bU_{\mathrm{rf}}(\bm{\theta}),\bU_{\mathrm{bb}}),\label{Urftheta}\\
&\mathcal{Y}\triangleq \{1,2,\ldots,N_rN_{\mathrm{rf}}\}.
\end{align}
Note that $\bU_{\mathrm{rf}}(\bm{\theta})$ in \eqref{Urftheta} means that $\bU_{\mathrm{rf}}$ is a function of $\bm{\theta}$, which can be written element-wisely as:
\begin{equation}
\bU_{\mathrm{rf}}(i,j) = e^{\sqrt{-1}\bm{\theta}((j-1)N_r+i)},\forall i\in\mathcal{N}_r, j\in\mathcal{N}_{\mathrm{rf}},
\end{equation}
where $\sqrt{-1}$ denotes the imaginary unit.

With the above derivation, we transform the intractable constant modulus constraints into linear constraints equivalently. From problem $\mathcal{P}4$, we can see that the objective function is non-convex. To address this issue, according to the technique of SCA, we need to find a surrogate function of $f(\bm{\theta})$ first. Let $r$ and $\bm{\theta}_r$ denote the iteration number and the current point in the $r$th iteration. Under the given local point $\bm{\theta}_r$, we can obtain one of the corresponding surrogate functions denoted by $\hat{f}(\bm{\theta},\bm{\theta}_r)$ via exploiting the first-order Taylor approximation, which is given by
\begin{align}
&\hat{f}(\bm{\theta},\bm{\theta}_r) = f(\bm{\theta}_{r}) + \bm{\gamma}^H_{\bm{\theta}_r}(\bm{\theta}-\bm{\theta}_r) + \tau\|\bm{\theta}-\bm{\theta}_{r}\|^2,\label{objappr}\\
&\bm{\gamma}_{\bm{\theta}_r} = \bigtriangledown_{\bm{\theta}} f(\bm{\theta})|_{\bm{\theta}=\bm{\theta}_r}=-\text{vec}\{2\mathbb{R}[\sqrt{-1}\bU_{\mathrm{rf},r}^{*}\circ\bF_r]\},\label{gradient1}\\
&\bF_r = \left(\sum\limits_{k=1}^K\bH_k\bV_k\bV_k^H\bH_k^H+\sigma^2\bI\right)\bU_{\mathrm{rf},r}\bU_{\mathrm{bb}}\bU_{\mathrm{bb}}^H-\sum\limits_{k=1}^K\bH_k\bV_k\bU_{\mathrm{bb}}^H,\label{gradient2}\\
&\bU_{\mathrm{rf},r}(i,j) = e^{\sqrt{-1}\bm{\theta}_r((j-1)N_{r}+i)},\forall i\in\mathcal{N}_r, j\in\mathcal{N}_{\mathrm{rf}}.\label{gradient3}
\end{align}
Note that the third term in \eqref{objappr} is a proximal regularization term with $\tau>0$ being a small positive number to guarantee the strong convexity and to control the convergence rate \cite{Liu2019}. $\bm{\gamma}_{\bm{\theta}_r}$ is the gradient of $f(\bm{\theta})$ with respect to $\bm{\theta}$ at point $\bm{\theta}_r$, which is calculated by the chain rule. With the above derivation, we can see that $\hat{f}(\bm{\theta},\bm{\theta}_r)$ is the upper bound of $f(\bm{\theta}_{r})$. Besides, $\hat{f}(\bm{\theta},\bm{\theta}_r)$ and $f(\bm{\theta}_{r})$ have the same values and gradient at point $\bm{\theta}_r$.
Thus, $\hat{f}(\bm{\theta},\bm{\theta}_r)$ is a valid surrogate function at point $\bm{\theta}_r$ \cite{Boyd2004} and the issue of non-convexity of $f(\bm{\theta})$ is addressed.

According to the procedure of SCA, we can update $\bm{\theta}$ and $\bU_{\mathrm{rf}}$ by solving the following approximated problem of $\mathcal{P}4$:
\begin{equation}\label{Prothetaapp}
\begin{split}
\mathcal{P}5\!:\;&\opmin_{\bm{\theta}} \hat{f}(\bm{\theta},\bm{\theta}_r)\\
&\text{subject to} ~~ -\pi\leq \bm{\theta}(i)\leq \pi, \forall i\in\mathcal{Y}.
\end{split}
\end{equation}
Since problem $\mathcal{P}5$ is convex with respect to $\bm{\theta}$, we can optimize $\bm{\theta}$ by checking the first-order optimality condition, for which the optimal solution is given by
\begin{equation}\label{thetaupdate}
\bm{\theta}_{r+1}(i) = \text{mod}\left(\bm{\theta}_r(i) - \frac{\bm{\gamma}_{\bm{\theta}_r}(i)}{2\tau},2\pi\right), \forall i\in\mathcal{Y}.
\end{equation}
Finally, the updated variable $\bU_{\mathrm{rf},r+1}$ can be obtained by
\begin{equation}\label{urfupdate}
\bU_{\mathrm{rf},r+1}(i,j) = e^{\sqrt{-1}\bm{\theta}_{r+1}((j-1)N_r+i)}, \forall i\in\mathcal{N}_r,j\in\mathcal{N}_{\mathrm{rf}}.
\end{equation}

The SCA-based algorithm to address problem $\mathcal{P}3$ is summarized in Algorithm \ref{SCAmethod}. According to the analysis in \cite{Liu2019} and \cite{Razaviyayn2013}, the proposed SCA-based algorithm can guarantee the convergence of a local optimum theoretically when $\tau$ is chosen properly. However, since the surrogate function is chosen based on the Taylor expansion which does not fully exploit the special structure of problem $\mathcal{P}3$. Hence, it may lead to high computational complexity and slow convergence rate. In the following, we propose an alternative effective approach with lower complexity in handling \eqref{ProUrfli}.

\begin{algorithm}[t]

\caption{\label{SCAmethod}The SCA-based Algorithm for Solving Problem $\mathcal{P}3$}

\textbf{Set }{$r=0$, $\tau>0$, and $\epsilon>0$}{\small\par}

\textbf{Repeat}

\textbf{\,\,\,\,\,Step 1: }{Calculate $\bm{\gamma}_{\bm{\theta}_r}$ according to \eqref{gradient1}, \eqref{gradient2}, and \eqref{gradient3};}

\textbf{\,\,\,\,\,Step 2: }{Updata $\bm{\theta}_{r+1}$ according to \eqref{thetaupdate};}

\textbf{\,\,\,\,\,Step 3: }{Update $\bU_{\mathrm{rf},r+1}$ according to \eqref{urfupdate};}

\textbf{\,\,\,\,\,Step 4: }{$r=r+1$;}

\textbf{until }the decrease of the objective function in problem $\mathcal{P}3$ is less than $\epsilon$.
\end{algorithm}

\subsubsection{Low-complexity Design via BCD}
Considering the tradeoff between the performance and complexity, we develop an alternative low-complexity algorithm to address problem $\mathcal{P}3$ by exploiting BCD. Based on further manipulation, problem $\mathcal{P}3$ can be equivalently converted as:
\begin{equation}\label{ProUrfqua}
\begin{split}
\mathcal{P}6\!:\;&\opmin_{\bU_{\mathrm{rf}}} \trace(\bU_{\mathrm{rf}}^H\bA\bU_{\mathrm{rf}}\bC)-2\mathbb{R}\{\trace(\bU_{\mathrm{rf}}^H\bB)\}\\
&\text{subject to} ~~|\bU_{\mathrm{rf}}(i,j)|=1, \forall i\in\mathcal{N}_{r}, j\in\mathcal{N}_{\mathrm{rf}},
\end{split}
\end{equation}
where $\bA\triangleq \sum\limits_{k=1}^K\bH_k\bV_k\bV_k^H\bH_k^H+\sigma^2\bI$, $\bB\triangleq \sum\limits_{k=1}^K \bH_k\bV_k\bU_{\mathrm{bb}}^H$, and $\bC\triangleq \bU_{\mathrm{bb}}\bU_{\mathrm{bb}}^H$. Since the unit modulus constraints are separable, inspired by \cite{ShiM2018}, we can update $\bU_{\mathrm{rf}}$ by applying the BCD type algorithm, i.e., in each step we only update one entry of $\bU_{\mathrm{rf}}$ by fixing others. Without loss of generality, by defining
\begin{equation}
\phi(\bU_{\mathrm{rf}}) = \trace(\bU_{\mathrm{rf}}^H\bA\bU_{\mathrm{rf}}\bC)-2\mathbb{R}\{\trace(\bU_{\mathrm{rf}}^H\bB)\},
\end{equation}
we investigate the problem of minimizing $\phi(\bU_{\mathrm{rf}})$ with respect to $\bU_{\mathrm{rf}}(i,j)$ for a particular $i\in\mathcal{N}_r$ and $j\in\mathcal{N}_{\mathrm{rf}}$ subject to the unit modulus constraint $|\bU_{\mathrm{rf}}(i,j)|=1$, i.e.,
\begin{equation}\label{APP2}
\mathcal{P}7\!:\;\opmin_{|\bU_{\mathrm{rf}}(i,j)|=1}~\phi(\bU_{\mathrm{rf}}).
\end{equation}
It can be observed that the objective function $\phi(\bU_{\mathrm{rf}})$ can be re-expressed as a quadratic function with respect to $\bU_{\mathrm{rf}}(i,j)$, i.e., $\tilde{\phi}(\bU_{\mathrm{rf}}(i,j))\triangleq a|\bU_{\mathrm{rf}}(i,j)|^2-2\mathbb{R}\{b^{*}\cdot\bU_{\mathrm{rf}}(i,j)\}$ for some real number $a$ and some complex number $b$ that will be explained later. Due to the unit modulus constraint, the first term of $\tilde{\phi}(\bU_{\mathrm{rf}}(i,j))$ is a constant. Then problem $\mathcal{P}7$ can be simplified as
\begin{equation}\label{APP3}
\mathcal{P}8\!:\;\opmax_{|\bU_{\mathrm{rf}}(i,j)|=1}~\mathbb{R}\{b^{*}\cdot\bU_{\mathrm{rf}}(i,j)\}.
\end{equation}
It is clear that the optimal solution of $\bU_{\mathrm{rf}}(i,j)$ to problem $\mathcal{P}8$ is equal to $\frac{b}{|b|}$. Hence, we only need to obtain $b$ for the update of $\bU_{\mathrm{rf}}(i,j)$.

Now we propose a handy method to update the complex number $b$. First, the following equality holds \cite{Petersen2012}:
\begin{equation}\label{der1}
\left.\bigtriangledown_{\bU_{\mathrm{rf}}^{*}(i,j)}\tilde{\phi}(\bU_{\mathrm{rf}}(i,j))\right|_{\bU_{\mathrm{rf}}(i,j)=\tilde{\bU}_{\mathrm{rf}}(i,j)}=\frac{1}{2}\left(a\tilde{\bU}_{\mathrm{rf}}(i,j)-b\right).
\end{equation}
Besides, we have \cite{Petersen2012}
\begin{equation}\label{der2}
\left.\bigtriangledown_{\bU_{\mathrm{rf}}^{*}}\tilde{\phi}(\bU_{\mathrm{rf}})\right|_{\bU_{\mathrm{rf}}=\tilde{\bU}_{\mathrm{rf}}}=\frac{1}{2}\left(\bA\tilde{\bU}_{\mathrm{rf}}\bC-\bB\right).
\end{equation}
Combining \eqref{der1} and \eqref{der2}, we have $[\bA\tilde{\bU}_{\mathrm{rf}}\bC-\bB]_{ij}=a\tilde{\bU}_{\mathrm{rf}}(i,j)-b$. By expanding $[\bA\tilde{\bU}_{\mathrm{rf}}\bC]_{ij}$ and checking the coefficient of $\tilde{\bU}_{\mathrm{rf}}(i,j)$, we have
\begin{equation}
a\tilde{\bU}_{\mathrm{rf}}(i,j)=\bA(i,i)\tilde{\bU}_{\mathrm{rf}}(i,j)\bC(j,j).
\end{equation}
Hence, $b$ can be updated according to the following equation:
\begin{equation}
b=\bA(i,i)\tilde{\bU}_{\mathrm{rf}}(i,j)\bC(j,j)-[\bA\tilde{\bU}_{\mathrm{rf}}\bC]_{ij}+\bB(i,j).
\end{equation}

Considering the above analysis, we can update the entries of $\bU_{\mathrm{rf}}$ iteratively. The corresponding algorithm for solving problem $\mathcal{P}6$ is summarized in Algorithm \ref{Appendix}, where we need to accordingly update $\bQ$ in Step 3 once $\bU_{\mathrm{rf}}(i,j)$ is updated (which is done in Step 4). As we can see, Step 3 is the most costly step requiring complexity $\mathcal{O}(N_rN_{\mathrm{rf}})$. Hence, it can be shown that the algorithm has complexity of $\mathcal{O}(N_r^2N_{\mathrm{rf}}^2)$.

From the above derivation, it is shown that we can obtain the optimal solution of each subproblem for one element of the receive analog beamforming while fixing the others. Considering the concept of the BCD algorithm \cite{Bertsekas2016}, the proposed algorithm in Algorithm \ref{Appendix} can converge to a stationary point of problem $\mathcal{P}6$. Furthermore, as compared with the SCA-based algorithm, this element-wise update in the BCD method can reduce the computational complexity by exploiting the special structures of the constant modulus constraints at the expense of certain performance degradation, which will be discussed in Section \uppercase\expandafter{\romannumeral3}-D.
\begin{algorithm}[t]

\caption{\label{Appendix}BCD-type Algorithm for Solving Problem $\mathcal{P}6$}

\textbf{Set }{$r=0$, $\bQ_{r}=\bA\bU_{\mathrm{rf},r}\bC$, and $\epsilon>0$}{\small\par}

\textbf{Repeat}

\textbf{\,\,\,\,\,For }{$i\in\mathcal{N}_r$ and $j\in\mathcal{N}_{\mathrm{rf}}$}

\textbf{\,\,\,\,\,\,\,\,\,\,Step 1: }{Calculate $b=\bA(i,i)\bU_{\mathrm{rf},r}(i,j)\bC(j,j)-\bQ_{r}(i,j)+\bB(i,j)$;}

\textbf{\,\,\,\,\,\,\,\,\,\,Step 2: }{Calculate $x=\frac{b}{|b|}$;}

\textbf{\,\,\,\,\,\,\,\,\,\,Step 3: }{Update $\bQ_{r+1}=\bQ_r+(x-\bU_{\mathrm{rf},r}(i,j))\bA(:,i)\bC(j,:)$;}

\textbf{\,\,\,\,\,\,\,\,\,\,Step 4: }{Update $\bU_{\mathrm{rf},r+1}(i,j)=x$;}

\textbf{\,\,\,\,\,end}

\textbf{\,\,\,\,\,$r=r+1$;}

\textbf{until }the decrease of the objective function in problem $\mathcal{P}6$ is less than $\epsilon$.
\end{algorithm}

\subsection{Optimization of Receive Digital Beamforming $\bU_{\mathrm{bb}}$}
Then, we optimize $\bU_{\mathrm{bb}}$ under fixed $\bU_{\mathrm{rf}}$ and $\{\bV_{k}\}$, for which we need to solve an unconstrained convex optimization problem given as:
\begin{equation}
\mathcal{P}9\!:\;\opmin_{\bU_{\mathrm{bb}}}~\text{MSE}(\{\bV_k\},\bU_{\mathrm{rf}},\bU_{\mathrm{bb}}).
\end{equation}
The receive digital beamforming to problem $\mathcal{P}9$ can be updated by applying the first-order optimality condition, which is given by
\begin{equation}\label{leastsqsol}
\bU_{bb}^{\mathrm{opt}} = \left(\bU_{\mathrm{rf}}^H\left(\sum\limits_{k=1}^K\bH_k\bV_k\bV_k^H\bH_k^H+\sigma^2\bI\right)\bU_{\mathrm{rf}}\right)^{-1}\bU_{\mathrm{rf}}^H\left(\sum\limits_{k=1}^K\bH_k\bV_k\right).
\end{equation}
From \eqref{leastsqsol}, it can be observed that the expression of $\bU_{\mathrm{bb}}$ has a sum-MMSE structure, which is different form the convectional MMSE receiver for multiuser massive MIMO communication systems in the form of $\left(\bU_{\mathrm{rf}}^H\left(\sum\limits_{k=1}^K\bH_k\bV_k\bV_k^H\bH_k^H+\sigma^2\bI\right)\bU_{\mathrm{rf}}\right)^{-1}\bU_{\mathrm{rf}}^H\bH_k\bV_k$ for estimating the individual message $\bs_k$ from device $k$ \cite{Krishnan2014}. More specifically, for the term outside the matrix inversion, we have $\bU_{\mathrm{rf}}^H\left(\sum\limits_{k=1}^K\bH_k\bV_k\right)$ in \eqref{leastsqsol} for estimating $\bs$ in AirComp but $\bU_{\mathrm{rf}}^H\bH_k\bV_k$ in conventional MMSE receiver for individually detecting $\bs_k$'s in communications. This is due to the fact that the signals from all the devices are exploited concurrently and beneficially to assist functional computation in massive MIMO AirComp systems, which is in shape contrast to the conventional multi-user massive MIMO communication systems by treating signals from different devices as harmful inter-device interference.

\subsection{Overall Algorithms}
According to the aforementioned results, the proposed hybrid beamforming designs for massive MIMO AirComp systems, named Lagrange-SCA and Lagrange-BCD, are summarized in Algorithm \ref{AirComp-MO}.
\begin{algorithm}[t]

\caption{\label{AirComp-MO}Pseudo-code of Proposed Hybrid Beamforming with SCA/BCD}

\textbf{Initialize }{$\bV_k$, $\bU_{\mathrm{rf}}$, and $\bU_{\mathrm{bb}}$, such that they meet all the constraints;}{\small\par}

\textbf{Repeat}

\textbf{\,\,\,\,\,\,\,\,\,\,Step 1: }{Optimize $\bV_k, \forall k\in\mathcal{K}$, using the Lagrange duality method;}

\textbf{\,\,\,\,\,\,\,\,\,\,Step 2: }{Optimize $\bU_{\mathrm{rf}}$ using SCA or BCD;}

\textbf{\,\,\,\,\,\,\,\,\,\,Step 3: }{Optimize $\bU_{\mathrm{bb}}$ according to \eqref{leastsqsol};}

\textbf{until }a stopping criterion is satisfied.
\end{algorithm}

Now, we investigate the complexity of Algorithm \ref{AirComp-MO} for designing the hybrid beamforming in the massive MIMO AirComp system, where only the dominant computational complexity with respect to $N_r$ is considered. In Step 1 of Algorithm \ref{AirComp-MO}, the bisection method for solving $\mu_k$ requires a complexity independent of $N_r$. Then, the complexity in calculating the matrix inverse in \eqref{Vkoptli} of Step 1 is $\mathcal{O}(KN_tN_{\mathrm{rf}}N_r)$. Similarly, the dominant computational complexity of the SCA-based algorithm is caused by the calculation of the gradient of the objective function, which can be expressed by $\mathcal{O}(N_r^2K(L+N_{\mathrm{rf}}))$. Also, the complexity of the BCD-based algorithm is $\mathcal{O}(N_r^2N_{\mathrm{rf}}^2)$ \cite{ShiM2018}. Since the number of wireless devices can be large, the complexity of the Lagrange-SCA algorithm is generally higher than that of the Lagrange-BCD algorithm. Finally, the complexity of optimization of $\bU_{\mathrm{bb}}$ is $\mathcal{O}(N_r^2(N_t+N_{\mathrm{rf}}))$.

Furthermore, the convergence of the proposed algorithms in Algorithm \ref{AirComp-MO} is obtained in the following theorem.

\theorem Any limiting point of the sequence generated by the Lagrange-SCA algorithm or the Lagrange-BCD algorithm in Algorithm \ref{AirComp-MO} is a stationary point of problem $\mathcal{P}1$.

\begin{IEEEproof}
See Appendix B.
\end{IEEEproof}


Considering Theorem 1, the objective function values of $\mathcal{P}1$ generated by the Lagrange-SCA and the Lagrange-BCD algorithms both decrease monotonically with respect to the number of iterations, for which the convergence speed will be validated in Section \uppercase\expandafter{\romannumeral5}.

\section{Analysis of massive MIMO AirComp systems}
In this section, we consider the special case with a fully-digital receiver (i.e., $N_{\mathrm{rf}}=N_r$ and $\bU_{\mathrm{rf}}=\bI$) to gain more design insights. This generally serves as an performance upper bound for other cases with $N_{\mathrm{rf}} < N_r$. For ease of analysis, we first focus on the case with a fixed transmit beamforming given by
\begin{equation}\label{noCSI}
\bV_k^H\bV_k=\frac{P}{L}\bI.
\end{equation}
Moreover, since we consider massive MIMO AirComp systems, we impose the following assumption.
\assump When $N_r$ is sufficiently large, the channel matrices between different wireless devices and the AP is asymptotically orthogonal, i.e.,
\begin{align}
\bH_k^H\bH_{k^{'}} &\approx \bm{0}, \forall k,k^{'} \in \mathcal{K}, k\neq k^{'},\label{Orthog}\\
\bH_k^H\bH_{k} &\approx \beta N_{r}\bI, \forall k\in \mathcal{K},\label{Orthog1}
\end{align}
where $\beta$ denotes the path loss from device to the AP.

This assumption is reasonable due to the properties of the massive MIMO technologies \cite{Marzetta2010}. In the following, we analyze the computation MSE performance and the corresponding receive digital beamforming design for the case with $N_r$ being sufficiently large. First, we derive the optimal fully-digital sum-MMSE receiver by setting $\bU_{\mathrm{rf}}=\bI$ in \eqref{leastsqsol}, which is given by
\begin{equation}\label{SZF}
\bU_{\mathrm{smmse}}=\left(\sum\limits_{k=1}^K\bH_k\bV_k\bV_k^H\bH_k^H+\sigma^2\bI\right)^{-1}\left(\sum\limits_{k=1}^K\bH_k\bV_k\right).
\end{equation}

With the help of \eqref{SZF} and the aforementioned assumptions, then we demonstrate how massive MIMO technologies affect the performance of the AirComp systems and how the receiver in \eqref{SZF} can be simplified via the following theorem and lemma.

\theorem When $N_r$ is sufficiently large, the computation MSE of massive MIMO AirComp systems, given by
\begin{equation}\label{digitalMSE}
\text{MSE}(\{\bV_k\},\bU_{\mathrm{bb}}) = \sum\limits_{k=1}^K\trace[(\bU_{\mathrm{bb}}^H\bH_k\bV_k-\bI)(\bU_{\mathrm{bb}}^H\bH_k\bV_k-\bI)^H]+\sigma^2\trace(\bU_{\mathrm{bb}}^H\bU_{\mathrm{bb}}),
\end{equation}
is inversely proportional to $N_r$ and can be written as follows:
\begin{equation}\label{MSEapprx}
\text{MSE}(\{\bV_k\},\bU_{\mathrm{bb}}) \approx \frac{KL^2\sigma^2}{\beta N_rP}.
\end{equation}
In particular, as $N_r\rightarrow \infty$, we have $\text{MSE}(\{\bV_k\},\bU_{\mathrm{bb}})\rightarrow 0$.

\begin{IEEEproof}
See Appendix C.
\end{IEEEproof}

\lemma When $N_r$ is sufficiently large, the optimal sum-MMSE receiver in \eqref{SZF} can be approximated as
\begin{equation}\label{simpu1}
\tilde{\bU}_{\mathrm{smmse}}=\sum\limits_{k=1}^K\bH_k\bV_k(\sigma^2\bI+\beta N_r\bV_k^H\bV_k)^{-1}.
\end{equation}
Under the transmit beamforming $\{\bV_k\}$ with $\bV_k^H\bV_k=\frac{P}{L}$ in \eqref{noCSI}, $\tilde{\bU}_{\mathrm{smmse}}$ in \eqref{simpu1} is rewritten as
\begin{equation}
\tilde{\bU}_{\mathrm{smmse}} = \sum\limits_{k=1}^K\frac{L}{\beta N_rP}\bH_k\bV_k.
\end{equation}

\begin{IEEEproof}
See Appendix D.
\end{IEEEproof}

From Theorem 2, we know that $\text{MSE}(\{\bV_k\},\bU_{\mathrm{bb}})$ increases with $K$, which indicates that the computation MSE performance is seriously degraded as the number of devices increases. To combat the vanishing computation MSE performance due to large $K$, we have two options: increasing the transmit power at the wireless devices or increasing the number of receive antennas at the AP. As such, the exploitation of massive MIMO techniques in AirComp systems is a practical solution since the former option is not energy-efficient. It is also observed from Theorem 2 that the computation MSE is not sensitive to $L$ due to our presumption that $L\leq\min(N_{\mathrm{rf}},N_t)\ll N_r$. Besides, exploiting Lemma 2, the sum-MMSE receiver can be significantly simplified under the considered special case with large $N_r$

\begin{figure}[htbp]
\centering
\includegraphics[width=4.5in]{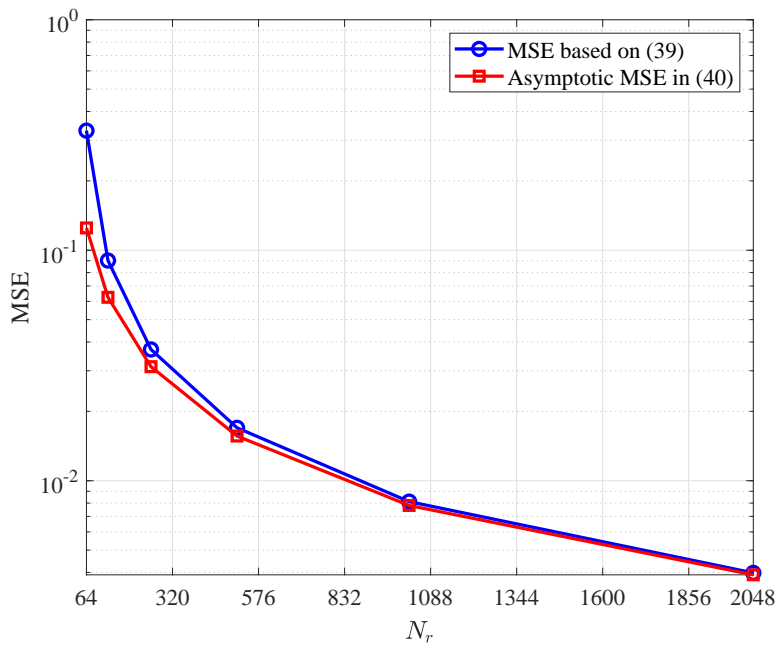}
\caption{The computation MSE performance of \eqref{digitalMSE} and \eqref{MSEapprx} versus $N_r$ when $N_t=2$, $K=20$, $L=2$, and $\text{SNR}=10\;\text{dB}$.}
\label{PVT}
\end{figure}

For illustration, in Fig. \ref{PVT}, we verify the accuracy of the derived asymptotic MSE in \eqref{MSEapprx} by comparing with the practical one obtained based on \eqref{digitalMSE}, where we set $\text{SNR}=\frac{\beta P}{\sigma^2}=10\;\text{dB}$, $\sigma=1$, $N_t=2$, $K=20$, and $L=2$. It can be seen in Fig. \ref{PVT} that the gap between the asymptotic MSE and the practical one decreases as $N_r$ increases. Specifically, the gap becomes negligible when $N_r\geq 512$.

\section{Simulation Results}
In this section, we evaluate the computation MSE performance of the proposed hybrid beamforming design approaches for massive MIMO AirComp systems, as compared with following benchmark schemes.
\begin{itemize}
\item\; FD-ZF: For this scheme, inspired by \cite{ZhuAir1}, a fully-digital orthogonal receive beamforming is adopted at the AP. Then, we alternately optimize the transmit beamforming and the receive beamforming. Specifically, the transmit beamforming is updated by the ZF method, i.e., forcing the first term of $\eqref{MSETRUE}$ to be zero. Second, by tightening the power constraints, an approximated problem of the fully-digital receive beamforming is formulated and solved by exploiting differential geometry \cite{Toponogov2006};
\item\; FD: For this scheme, we alternately optimize the transmit beamforming and the fully-digital receive beamforming by exploiting the Lagrange duality method and the sum-MMSE receiver in \eqref{SZF}, respectively.
\end{itemize}

Among the simulation experiments, each channel is assumed to be normalized i.i.d. Rayleigh fading. The initial phases of the receive analog beamforming, $\bU_{\mathrm{rf}}$, follow an uniform distribution over $[-\pi,\pi]$. Also, we set $N_t=L$, $\tau=0.2$, and $\epsilon=10^{-3}$. Besides, all simulation results are averaged over $500$ channel realizations.

\begin{figure}[htbp]
\centering
\includegraphics[width=4.5in]{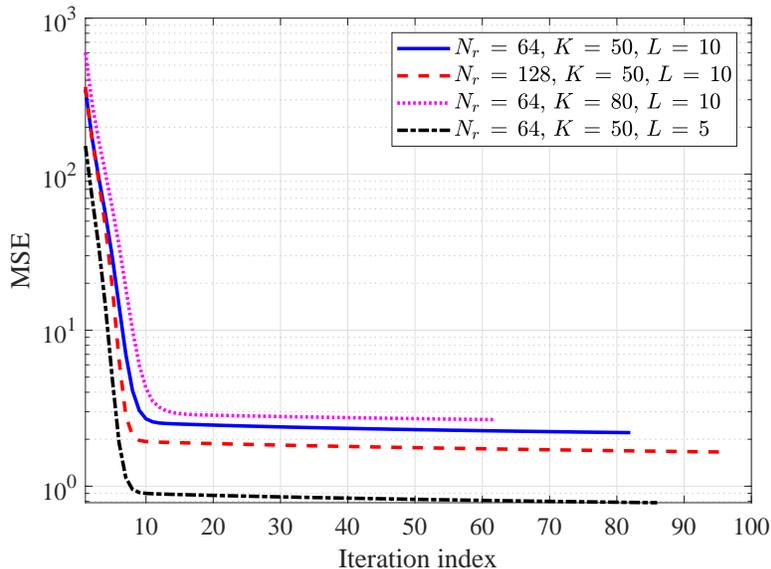}
\caption{The convergence performance of the Lagrange-SCA algorithm with $N_{\mathrm{rf}}=16$ and $\text{SNR}=10\;\text{dB}$.}
\label{fig:fig3}
\end{figure}

\begin{figure}[htbp]
\centering
\includegraphics[width=4.5in]{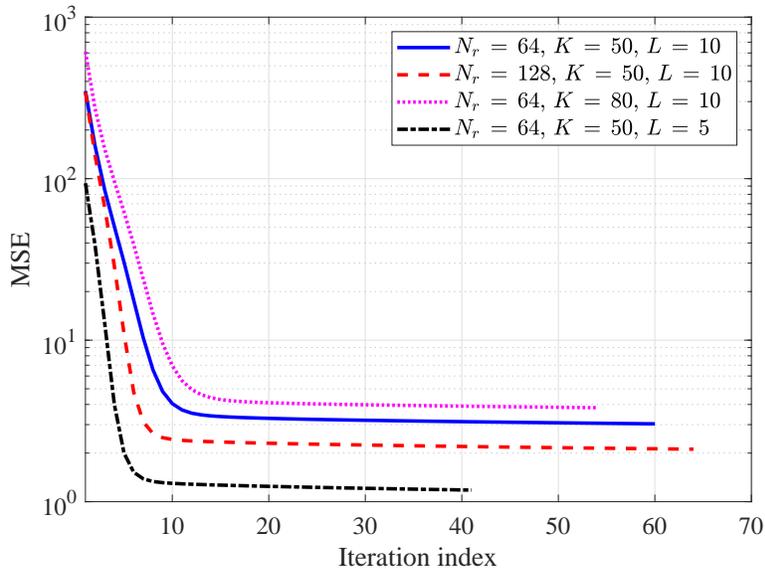}
\caption{The convergence performance of the Lagrange-BCD algorithm with $N_{\mathrm{rf}}=16$ and $\text{SNR}=10\;\text{dB}$.}
\label{fig:fig31}
\end{figure}

Fig. \ref{fig:fig3} and Fig. \ref{fig:fig31} illustrate the convergence behavior for the Lagrange-SCA and the Lagrange-BCD algorithms in Algorithm \ref{AirComp-MO}, where $N_{\mathrm{rf}}=16$ and $\text{SNR}=10\;\text{dB}$, respectively. From these figures, we conclude that the proposed Lagrange-SCA algorithm and Lagrange-BCD algorithm both converge rapidly in a few iterations. Besides, it can be observed that the Lagrange-BCD algorithm always converges faster, while the Lagrange-SCA algorithm shows better performance. This unveils the tradeoff between the system performance and the computational complexity.

\begin{figure}[htbp]
\centering
\includegraphics[width=4.5in]{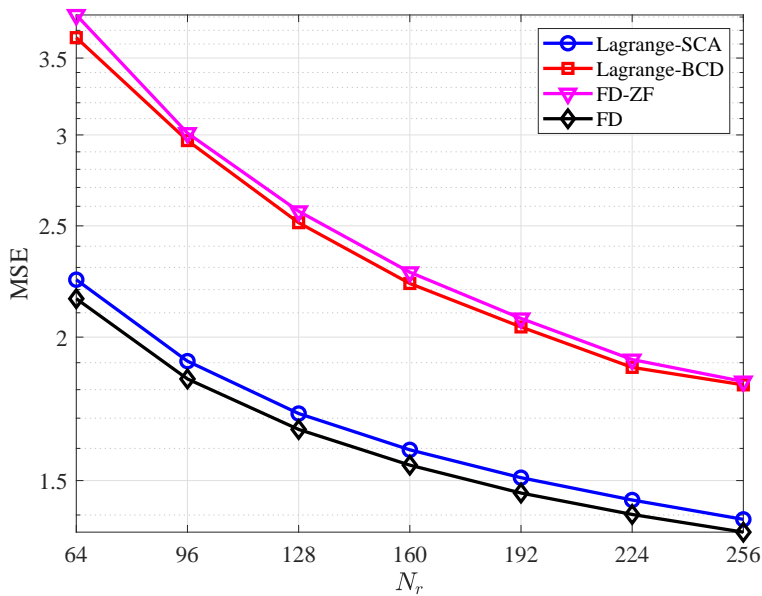}
\caption{The computation MSE performance versus $N_r$ when $N_t=10$, $N_{\mathrm{rf}}=10$, $K=50$, $L=10$, and $\text{SNR}=10\;\text{dB}$.}
\label{fig:fig5}
\end{figure}

\begin{figure}[htbp]
\centering
\includegraphics[width=4.5in]{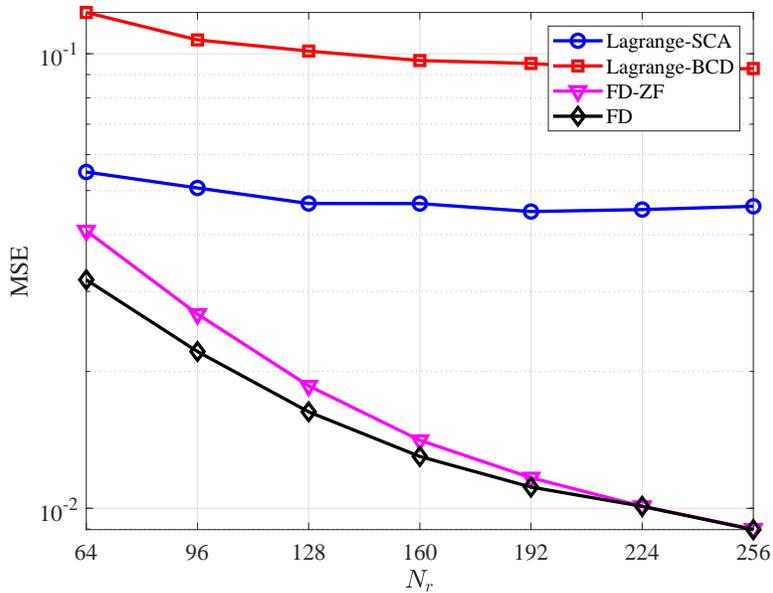}
\caption{The computation MSE performance versus $N_r$ when $N_t=1$, $N_{\mathrm{rf}}=1$, $K=20$, $L=1$, and $\text{SNR}=10\;\text{dB}$.}
\label{fig:fig5_single}
\end{figure}

Fig. \ref{fig:fig5} shows the computation MSE performance of multi-function/multi-modal ($L>1$) massive MIMO AirComp systems versus $N_r$ under different setups. From Fig. \ref{fig:fig5}, we can see that the computation MSE value of both proposed algorithms decreases considerably when the number of receive antennas increases, showing the effectiveness of applying massive MIMO. In particular, the proposed Lagrange-SCA and Lagrange-BCD algorithms, where hybrid receivers are adopted, outperform the FD-ZF algorithm with a fully-digital receiver. In fact, our proposed algorithms can exploit the structure of the beamforming problem, offering a better solution for massive MIMO AirComp systems than the heuristic algorithm based on the ZF method. It is also observed that the performance gaps between the FD-ZF algorithm and our proposed algorithms become smaller as $N_r$ increases. This is intuitive, as the FD-ZF can exploit the increased spatial degrees of freedom for enhancing the performance due to its fully-digital beamforming structure. Moreover, the computation MSE performance of the Lagrange-BCD algorithm and the FD-ZF algorithm coincides when $N_r=256$. It suggests that the diversity gain of the massive MIMO techniques can compensate for the performance degradation due to the drawbacks of the FD-ZF algorithm, which confirms the correctness of our analysis in Section \uppercase\expandafter{\romannumeral4}. Combining Figs. \ref{fig:fig3}, \ref{fig:fig31} and \ref{fig:fig5}, it is clear that the Lagrange-SCA algorithm outperforms the Lagrange-BCD algorithm, but at the cost of a slower convergence and a high complexity. Besides, the computation MSE performance of the proposed Lagrange-SCA algorithm is close to the performance upper bound achieved by the FD method, which verifies the effectiveness of the proposed hybrid beamforming design.


To unveil more insights, Fig. \ref{fig:fig5_single} shows MSE performance for the special case with single-function/single-modal with a single RF chain equipped at the AP (i.e., $L=1$ and $N_{\mathrm{rf}}=1$). It can be seen that the considered hybrid beamforming designs, including the Lagrange-SCA algorithm and the Lagrange-BCD algorithm, perform poorly in this case. It is mainly due to that the hybrid beamforming design reduces to the simple analog beamforming design when $N_{\mathrm{rf}}=1$, which can not exploit the spatial degrees of freedom and the array gain brought by massive MIMO.

\begin{figure}[htbp]
\centering
\includegraphics[width=4.5in]{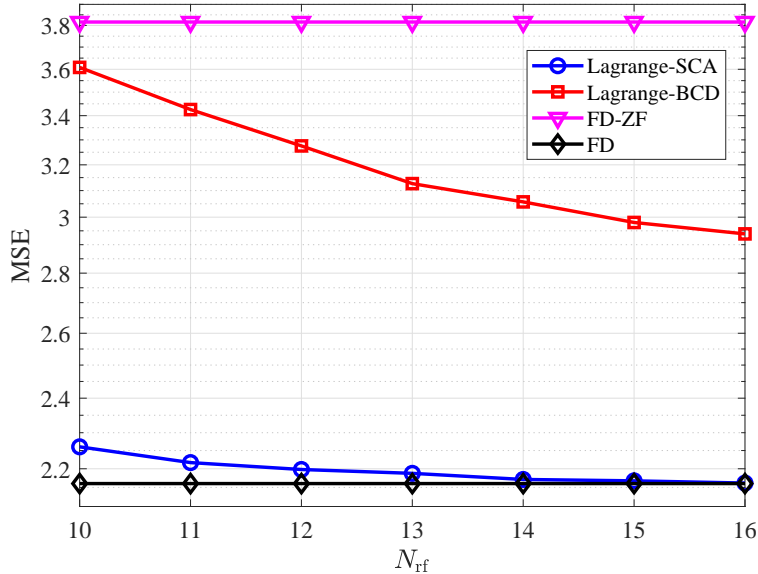}
\caption{The computation MSE performance versus $N_{\mathrm{rf}}$ when $N_r=64$, $N_t=10$, $K=50$, $L=10$, and $\text{SNR}=10\;\text{dB}$.}
\label{fig:figrf}
\end{figure}

Fig. \ref{fig:figrf} shows the computation MSE performance versus $N_{\mathrm{rf}}$ when $N_r=64$, $N_t=10$, $K=50$, $L=10$, and $\text{SNR}=10\;\text{dB}$. As a property of hybrid beamforming schemes, it can be observed that the MSE values achieved by the Lagrange-SCA and Lagrange-BCD algorithms reduce with increasing $N_{\mathrm{rf}}$. By adopting the minimum number of RF chains ($N_{\mathrm{rf}}\geq L$ and $L=10$), the Lagrange-SCA algorithm shows only slight performance degradation compared with the FD method while both algorithms achieve the same MSE when $N_{\mathrm{rf}}\geq 15$. Besides, the Lagrange-SCA and the Lagrange-BCD algorithms both outperform the FD-ZF algorithm even when $N_{\mathrm{rf}}=10$. This illustrates that our proposed algorithm can exploit the spatial degrees of freedom efficiently with much less RF chains. In contrast, the FD-ZF algorithm shows a constant computation MSE value when $N_{\mathrm{rf}}$ increases, as the performance of a fully-digital beamforming design is independent of the number of RF chains.
\begin{figure}[htbp]
\centering
\includegraphics[width=4.5in]{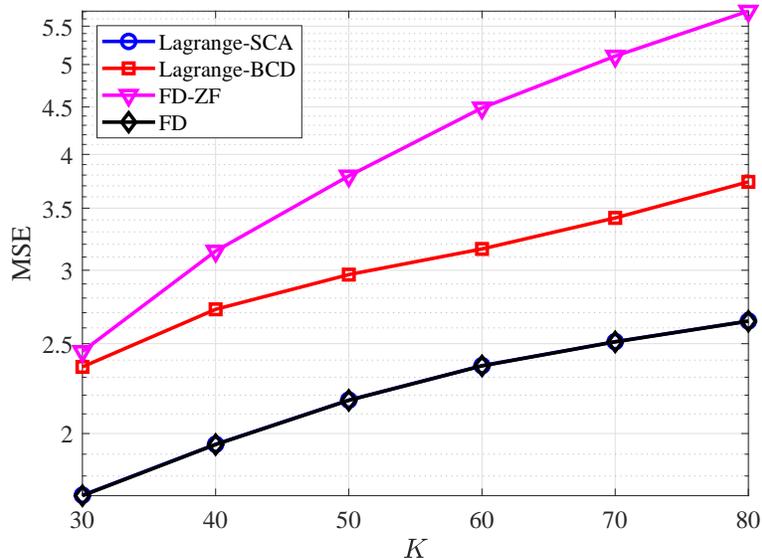}
\caption{The computation MSE performance versus $K$ when $N_r=64$, $N_t=10$, $N_{\mathrm{rf}}=16$, $L=10$, and $\text{SNR}=10\;\text{dB}$.}
\label{fig:fig6}
\end{figure}

The computation MSE performance versus $K$ is shown in Fig. \ref{fig:fig6} under $N_r=64$, $N_t=10$, $N_{\mathrm{rf}}=16$, $L=10$, and $\text{SNR}=10\;\text{dB}$. One can see that the MSE performance of all the considered schemes increases with $K$. This coincides with our discussion that supporting the connection of more wireless devices is at the cost of degrading computational accuracy which makes it more challenging to design a common receive fully-digital/hybrid beamforming to equalize all the wireless devices' channels. Besides, the performance of the Lagrange-SCA algorithm still approaches that of the benchmark FD method. Also, our proposed algorithms outperform the FD-ZF algorithm. Furthermore, the performance gap between our proposed algorithms and the FD-ZF algorithm increases as $K$ increases from $30$ to $80$. It is mainly due to the fact that the FD-ZF algorithm adopts an orthogonal receive beamforming which becomes a highly suboptimal solution jeopardizing the system performance when $K$ is large.

\begin{figure}[htbp]
\centering
\includegraphics[width=4.5in]{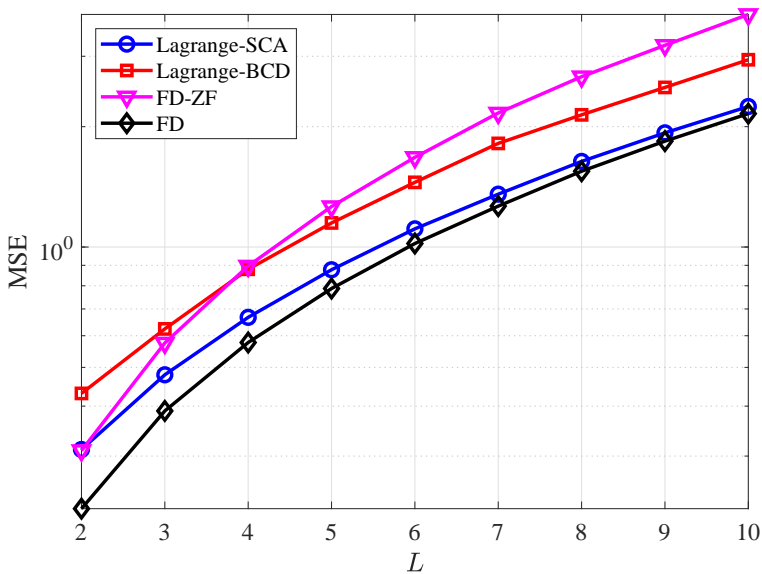}
\caption{The computation MSE performance versus $L$ when $N_r=64$, $N_{\mathrm{rf}}=10$, $K=50$, and $\text{SNR}=10\;\text{dB}$.}
\label{fig:fig7}
\end{figure}

Fig. \ref{fig:fig7} compares the MSE performance of different algorithms versus $L$ when $N_r=64$, $N_{\mathrm{rf}}=10$, $K=50$, and $\text{SNR}=10\;\text{dB}$. From this figure, the computation MSEs of all algorithms increase with the number of functions operated at the AP, implying that the multi-function operation at the AP leads to increased computation error. Besides, we can see that the FD-ZF algorithm shows less computation MSE than the Lagrange-BCD algorithm when $L\leq 3$, since the inter-function interference is insignificant and can be handled by the FD-ZF algorithm in such a case. In particular, the fully-digital beamforming structure in the FD-ZF algorithm shows its advantages of exploiting the spatial degrees and the array gain compared with the hybrid one in the Lagrange-BCD algorithm. However, these advantages are marginal as the inter-function interference becomes more severe in the regime of large $L$. This further verifies the effectiveness of our proposed algorithms in multi-function/multi-modal massive MIMO AirComp systems.

\begin{figure}[htbp]
\centering
\includegraphics[width=4.5in]{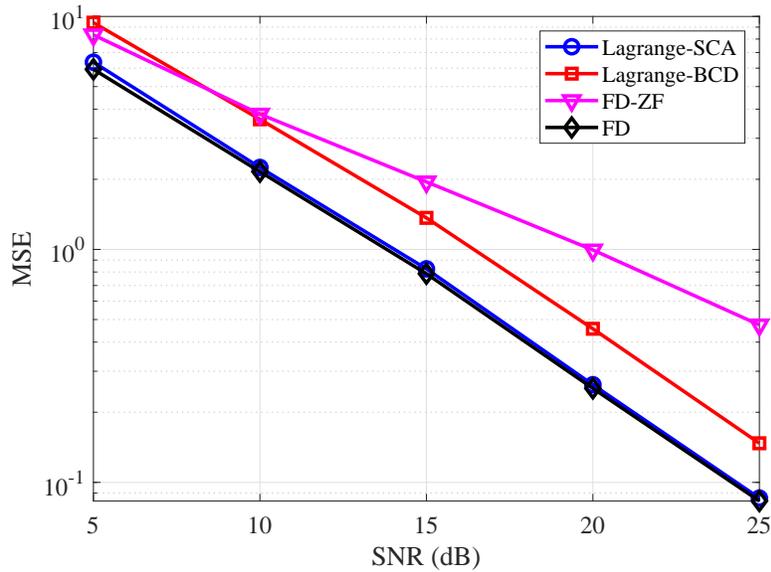}
\caption{The computation MSE performance versus $\text{SNR}$ when $N_r=64$, $N_t=10$, $N_{\mathrm{rf}}=10$, $K=50$, and $L=10$.}
\label{fig:fig8}
\end{figure}

Fig. \ref{fig:fig8} shows the MSE performance versus SNR under $N_r=64$, $N_t=10$, $N_{\mathrm{rf}}=10$, $K=50$, and $L=10$. We can see that the MSE values by all our considered schemes decreases monotonically as $\text{SNR}$ increases. Due to the advantages of the fully-digital beamforming structure, the FD-ZF algorithm achieves a smaller computation MSE than that of the Lagrange-BCD algorithm when SNR is less than $10\;\text{dB}$, while the opposite holds when SNR is larger than $10\;\text{dB}$. However, our proposed Lagrange-SCA algorithm outperforms the FD-ZF algorithm over the whole SNR regime due to the proposed resource optimization.

%

\section{conclusion}
In this paper, we exploited massive MIMO with hybrid beamforming for AirComp systems. We jointly optimized the transmit digital beamforming at devices, the receive analog and digital beamforming at the AP to minimize the computation MSE. To solve the non-convex hybrid beamforming problem for MSE minimization, we proposed alternating-optimization-based designs, in which we first optimized the transmit digital beamforming using the Lagrange duality method, then proposed two approaches to update the receive analog beamforming via SCA and BCD, respectively, and finally optimized the receive digital beamforming by applying the first-order optimality condition. To gain more insights, we analyzed the system performance for the special case with a fully-digital receive beamforming at AP and proved that the computation MSE is inversely proportional to the number of receive antennas for this special case. Our numerical results showed that the proposed algorithms achieve an outstanding performance that is close to the performance upper bound achieved under a fully-digital receiver but with a much smaller number of RF chains required.

\appendices{
\section{Proof of Lemma 1}
Let $\mu_k\geq 0$ denote the Lagrange multiplier associated with the power constraint for device $k$ in problem $\mathcal{P}2$. The Lagrangian of $\mathcal{P}2$ is denoted as:
\begin{equation}
\mathcal{L}(\bV_k)\triangleq \trace[(\bU_{\mathrm{bb}}^H\bU_{\mathrm{rf}}^H\bH_k\bV_k-\bI)(\bU_{\mathrm{bb}}^H\bU_{\mathrm{rf}}^H\bH_k\bV_k-\bI)^H] + \mu_k[\trace(\bV_k\bV_k^H)-P].
\end{equation}
According to the Karush-Kuhn-Tucker (KKT) conditions, the following equalities hold:
\begin{align}
&\bigtriangledown_{\bV_k}\mathcal{L}(\bV_k) = \mathbf{0},\label{KKT1}\\
&\trace(\bV_k\bV_k^H)-P\leq 0,\label{KKT2}\\
&\mu_k\geq 0,\label{KKT3}\\
&\mu_k(\trace(\bV_k\bV_k^H)-P)=0\label{KKT4}.
\end{align}
\eqref{KKT1} is the first-order optimality condition of $\mathcal{L}(\bV_k)$ with respect to $\bV_k$, which yields
\begin{equation}\label{Vkoptl1}
\bV_k = (\bH_k^H\bU_{\mathrm{rf}}\bU_{\mathrm{bb}}\bU_{\mathrm{bb}}^H\bU_{\mathrm{rf}}^H\bH_k+\mu_k\bI)^{-1}\bH_k^H\bU_{\mathrm{rf}}\bU_{\mathrm{bb}}.
\end{equation}

Therefore, once we obtain the optimal $\mu_k$, denoted by $\mu_k^{\mathrm{opt}}$, $\bV_k$ can be updated by substituting $\mu_k^{\mathrm{opt}}$ into \eqref{Vkoptl1}. Combining \eqref{KKT3} and \eqref{KKT4}, we optimize $\mu_k$ by considering two cases with $\mu_k=0$ and $\mu_k>0$. Let $\bV_k(\mu_k)$ denote the right-hand side of \eqref{Vkoptl1}. For the first case, if $\bH_k^H\bU_{\mathrm{rf}}\bU_{\mathrm{bb}}\bU_{\mathrm{bb}}^H\bU_{\mathrm{rf}}^H\bH_k$ is invertible and
\begin{equation}\label{muk0}
\trace[\bV_k(0)\bV_k^H(0)]< P,
\end{equation}
then the optimal transmit beamforming is given by $\bV_k(0)$. For the second case, the equality of constraint in problem $\mathcal{P}2$ holds according to \eqref{KKT4}.
\begin{equation}\label{Vkeq}
\trace[\bV_k(\mu_k)\bV_k^H(\mu_k)]= P.
\end{equation}

Substituting \eqref{Vkoptl1} into \eqref{Vkeq}, we have
\begin{equation}\label{Mubisection}
\trace[(\bH_k^H\bU_{\mathrm{rf}}\bU_{\mathrm{bb}}\bU_{\mathrm{bb}}^H\bU_{\mathrm{rf}}^H\bH_k+\mu_k\bI)^{-2}\bH_k^H\bU_{\mathrm{rf}}\bU_{\mathrm{bb}}\bU_{\mathrm{bb}}^H\bU_{\mathrm{rf}}^H\bH_k]=P.
\end{equation}
Note that $\mu_k$ must be positive in this case and the left-hand side is a decreasing function with respect to $\mu_k$ for $\mu_k> 0$. Then we can obtain the optimal Lagrange multiplier in \eqref{Mubisection} via a bisection search. Finally, the transmit beamforming can be updated by substituting $\mu_k^{\mathrm{opt}}$ into \eqref{Vkoptl1}.

\section{Proof of theorem 1}
We first prove the existence of at least one limiting point before stating that any limit point of the sequence generated by the proposed algorithms is a stationary solution. In this paper, the feasible set of each variable ($\{\bV_k\}$, $\bU_{\mathrm{rf}}$, and $\bU_{\mathrm{bb}}$) is compact, respectively. Then, problem $\mathcal{P}1$ over their Cartesian product set is bounded. Therefore, the sequence generated by Algorithm \ref{AirComp-MO} is compact and bounded. Since any compact and bounded sequence must have at least one limiting point, we can claim the existence of a limiting point of our proposed algorithms.

Then, let $r$ denote the iteration number. Clearly, given $\bU_{\mathrm{rf},r}$ and $\bU_{\mathrm{bb},r}$, the optimal solution for $\{\bV_{k,r}\}$ can be obtained by using the Lagrange duality method and the KKT conditions, which leads to
\begin{equation}
\text{MSE}(\{\bV_{k,r+1}\},\bU_{\mathrm{rf},r},\bU_{\mathrm{bb},r})\leq \text{MSE}(\{\bV_{k,r}\},\bU_{\mathrm{rf},r},\bU_{\mathrm{bb},r}).
\end{equation}
Considering the discussion in Section \uppercase\expandafter{\romannumeral3}-B, Algorithm \ref{SCAmethod} and Algorithm \ref{Appendix} both establish the local convergence to the stationary solutions of problem $\mathcal{P}3$ when $\{\bV_{k,r}\}$ and $\bU_{\mathrm{bb},r}$ are fixed. Then the following inequality holds:
\begin{equation}
\text{MSE}(\{\bV_{k,r}\},\bU_{\mathrm{rf},r+1},\bU_{\mathrm{bb},r})\leq \text{MSE}(\{\bV_{k,r}\},\bU_{\mathrm{rf},r},\bU_{\mathrm{bb},r}).
\end{equation}
Similarly, when we update $\bU_{\mathrm{bb}}$ by using \eqref{leastsqsol} with given $\{\bV_{k,r}\}$ and $\bU_{\mathrm{rf,r}}$, we have
\begin{equation}
\text{MSE}(\{\bV_{k,r}\},\bU_{\mathrm{rf},r},\bU_{\mathrm{bb},r+1})\leq \text{MSE}(\{\bV_{k,r}\},\bU_{\mathrm{rf},r},\bU_{\mathrm{bb},r}).
\end{equation}
Note that in each iteration of Algorithm \ref{AirComp-MO}, the objective function value is non-increasing and also lower bounded by zero. Hence, the convergence of Algorithm \ref{AirComp-MO} follows.

\section{Proof of Theorem 2}
Substituting the optimal sum-MMSE receiver in \eqref{SZF} into \eqref{digitalMSE}, we have
\begin{align}
&\text{MSE}(\{\bV_k\},\bU_{\mathrm{bb}})\nonumber\\
&=\sum\limits_{k=1}^K\trace[(\bU_{\mathrm{bb}}^H\bH_k\bV_k\bV_k^H\bH_k^H\bU_{\mathrm{bb}}-\bU_{\mathrm{bb}}^H\bH_k\bV_k-\bV_k^H\bH_k^H\bU_{\mathrm{bb}}+\bI)+\sigma^2\bU_{\mathrm{bb}}^H\bU_{\mathrm{bb}}]\nonumber\\
&=KL+\trace[\bU_{\mathrm{bb}}^H\left(\sum\limits_{k=1}^K\bH_k\bV_k\bV_k^H\bH_k^H+\sigma^2\bI\right)\bU_{\mathrm{bb}}-\sum\limits_{k=1}^K(\bU_{\mathrm{bb}}^H\bH_k\bV_k+\bV_k^H\bH_k^H\bU_{\mathrm{bb}})]\nonumber\\
&=KL-\trace\left[\left(\sum\limits_{k=1}^K\bV_k^H\bH_k^H\right)\left(\sum\limits_{k=1}^K\bH_k\bV_k\bV_k^H\bH_k^H+\sigma^2\bI\right)^{-1}\left(\sum\limits_{k=1}^K\bH_k\bV_k\right)\right].\label{MSEAPPA1}
\end{align}

As we can see that the term $\left(\sum\limits_{k=1}^K\bH_k\bV_k\bV_k^H\bH_k^H+\sigma^2\bI\right)^{-1}$ is intractable. To simplify the below derivations, let us define the following matrix sequence:
\begin{equation}\label{Morridef}
\bA_n = \sum\limits_{k=1}^n\bH_k\bV_k\bV_k^H\bH_k^H+\sigma^2\bI, \forall n\in \mathcal{K}.
\end{equation}
Then we have $\bA_n = \bA_{n-1} + \bH_n\bV_n\bV_n^H\bH_n^H$ and $\bA_K = \sum\limits_{k=1}^K\bH_k\bV_k\bV_k^H\bH_k^H+\sigma^2\bI$. According to the Kailath Variant identity \cite{Petersen2012}, the inverse of $\bA_n$ is given by
\begin{equation}\label{Aninver}
\bA_n^{-1} = \bA_{n-1}^{-1}-\bA_{n-1}^{-1}\bH_n\bV_n(\bI+\bV_n^H\bH_n^H\bA_{n-1}^{-1}\bH_n\bV_n)^{-1}\bV_n^H\bH_n^H\bA_{n-1}^{-1}.
\end{equation}
Letting $n$ be $1$ in \eqref{Morridef} and using \eqref{Aninver}, we have
\begin{equation}
\bA_1^{-1} =\frac{1}{\sigma^2}\left(\bI-\bH_1\bV_1\left(\sigma^2\bI+\bV_1^H\bH_1^H\bH_1\bV_1\right)^{-1}\bV_1^H\bH_1^H\right).
\end{equation}
Then we check the inverse of $\bA_2$
\begin{equation}\label{MOrrder}
\begin{split}
\bA_2^{-1}&=\bA_{1}^{-1}-\bA_{1}^{-1}\bH_2\bV_2(\bI+\bV_2^H\bH_2^H\bA_1^{-1}\bH_2\bV_2)^{-1}\bV_2^H\bH_2^H\bA_{1}^{-1}\\
&\approx\bA_{1}^{-1}-\frac{1}{\sigma^2}\bH_2\bV_2\left(\bI+\frac{1}{\sigma^2}\bV_2^H\bH_2^H\bH_2\bV_2\right)^{-1}\bV_2^H\bH_2^H\\
&=\frac{1}{\sigma^2}\left(\bI-\sum\limits_{n=1}^2\bH_n\bV_n(\sigma^2\bI+\bV_n^H\bH_n^H\bH_n\bV_n)^{-1}\bV_n^H\bH_n^H\right),
\end{split}
\end{equation}
where the approximation is due to Assumption 1 stated in \eqref{Orthog} and \eqref{Orthog1}. Hence, from \eqref{MOrrder}, we can obtain the inverse of $\bA_K$ as follows:
\begin{equation}\label{AKinver}
\bA_K^{-1} \approx \frac{1}{\sigma^2}\left(\bI-\sum\limits_{n=1}^K\bH_n\bV_n(\sigma^2\bI+\bV_n^H\bH_n^H\bH_n\bV_n)^{-1}\bV_n^H\bH_n^H\right).
\end{equation}

Substituting \eqref{AKinver} into \eqref{MSEAPPA1} yields
\begin{align}
&\text{MSE}(\{\bV_k\},\bU_{\mathrm{bb}})\nonumber\\
&=KL-\frac{1}{\sigma^2}\trace\left[\left(\sum\limits_{k=1}^K\bV_k^H\bH_k^H\right)\bA_K^{-1}\left(\sum\limits_{k=1}^K\bH_k\bV_k\right)\right]\nonumber\\
&=KL-\frac{1}{\sigma^2}\trace\left[\sum\limits_{k=1}^K(\bV_k^H\bH_k^H\bH_k\bV_k-\bV_k^H\bH_k^H\bH_k\bV_k(\sigma^2\bI+\bV_k^H\bH_k^H\bH_k\bV_k)^{-1}\bV_k^H\bH_k^H\bH_k\bV_k)\right]\nonumber\\
&\approx KL-\frac{1}{\sigma^2}\sum\limits_{k=1}^K\trace(\beta N_r\bV_k^H\bV_k-\beta^2N_r^2\bV_k^H\bV_k(\sigma^2\bI+\beta N_r\bV_k^H\bV_k)^{-1}\bV_k^H\bV_k)\nonumber\\
&=KL-\sum\limits_{k=1}^K\trace(\beta N_r\bV_k^H\bV_k(\sigma^2\bI+\beta N_r\bV_k^H\bV_k)^{-1})\nonumber\\
&=KL-\sum\limits_{k=1}^K\trace\left(\frac{\beta N_rP}{L}\left(\sigma^2+\frac{\beta N_rP}{L}\right)^{-1}\bI\right)\nonumber\\
&=KL-\frac{KL\beta N_rP}{L\sigma^2+\beta N_rP}\nonumber\\
&=\frac{KL^2\sigma^2}{L\sigma^2+\beta N_rP},\label{MSEAPP2}
\end{align}
where the approximation and the fourth equality are due to \eqref{Orthog1} and \eqref{noCSI}, respectively. When $N_r$ is sufficiently large, $\beta N_rP$ becomes the dominated term of denominator such that \eqref{MSEAPP2} is reduced to
\begin{equation}\label{zeroMSE}
\text{MSE}(\{\bV_k\},\bU_{\mathrm{bb}})\approx \frac{KL^2\sigma^2}{\beta N_rP}.
\end{equation}
Obviously, when $N_r\rightarrow \infty$, \eqref{zeroMSE} goes to zeros. The result follows immediately.

\section{Proof of Lemma 2}
According to \eqref{AKinver}, \eqref{SZF} can be rewritten as
\begin{equation}\label{ReAPP1}
\begin{split}
\tilde{\bU}_{\mathrm{smmse}} &=  \frac{1}{\sigma^2}\left(\bI-\sum\limits_{k=1}^K\bH_k\bV_k(\sigma^2\bI+\bV_k^H\bH_k^H\bH_k\bV_k)^{-1}\bV_k^H\bH_k^H\right)\left(\sum\limits_{k=1}^K\bH_k\bV_k\right)\\
& \approx \sum\limits_{k=1}^K\frac{1}{\sigma^2}\left(\bH_k\bV_k-\bH_k\bV_k(\sigma^2\bI+\bV_k^H\bH_k^H\bH_k\bV_k)^{-1}\bV_k^H\bH_k^H\bH_k\bV_k\right),\\
& = \sum\limits_{k=1}^K\frac{1}{\sigma^2}\bH_k\bV_k\left(\bI-(\sigma^2\bI+\bV_k^H\bH_k^H\bH_k\bV_k)^{-1}\bV_k^H\bH_k^H\bH_k\bV_k\right),\\
& = \sum\limits_{k=1}^K\bH_k\bV_k(\sigma^2\bI+\bV_k^H\bH_k^H\bH_k\bV_k)^{-1}\\
& \approx \sum\limits_{k=1}^K\bH_k\bV_k(\sigma^2\bI+\beta N_r\bV_k^H\bV_k)^{-1},
\end{split}
\end{equation}
where the approximations are due to \eqref{Orthog} and \eqref{Orthog1}, respectively. For the special case where $\bV_k^H\bV_k=\frac{P}{L}\bI$ and $N_r$ is sufficiently large, \eqref{ReAPP1} can be simplified as:
\begin{equation}
\begin{split}
\tilde{\bU}_{\mathrm{smmse}} &= \sum\limits_{k=1}^K\frac{L}{L\sigma^2+\beta N_rP}\bH_k\bV_k\\
&= \sum\limits_{k=1}^K\frac{L}{\beta N_rP}\bH_k\bV_k.
\end{split}
\end{equation}
The result follows immediately.

}


\end{document}